\documentclass{article}
\usepackage{spconf,amsmath,graphicx}

\usepackage{cite}
\usepackage{txfonts}
\usepackage{setspace}
\usepackage{bm}
\usepackage{multirow}
\usepackage{subfig}
\usepackage{comment}
\usepackage{amssymb}
\usepackage[dvipdfmx]{hyperref}
\usepackage{xcolor}
\usepackage{setspace}

\title{All for One and One for All:\\Improving Music Separation by Bridging Networks}

\name{Ryosuke Sawata$^\dagger$, Stefan Uhlich$^\ddagger$, Shusuke Takahashi$^\dagger$ and Yuki Mitsufuji$^\dagger$} 
\address{$^\dagger$Sony Corporation, Tokyo, Japan \hspace{1cm}
$^\ddagger$Sony Europe B.V., Stuttgart, Germany}

\begin{document}
\ninept
\maketitle
\begin{abstract}
This paper proposes several improvements for music separation with \emph{deep neural networks} (DNNs), namely a \emph{multi-domain loss} (MDL) and two \emph{combination schemes}.
First, by using MDL we take advantage of the frequency and time domain representation of audio signals.
Next, we utilize the relationship among instruments by jointly considering them. We do this on the one hand by modifying the network architecture and introducing a \emph{CrossNet} structure. On the other hand, we consider combinations of instrument estimates by using a new \emph{combination loss} (CL).
MDL and CL can easily be applied to many existing DNN-based separation methods as they are merely loss functions which are only used during training and do not affect the inference step.
Experimental results show that the performance of \emph{Open-Unmix} (UMX), a well-known and state-of-the-art open-source library for music separation, can be improved by utilizing our above schemes.
Our modifications of UMX are open-sourced together with this paper.
\end{abstract}
\vspace{-1mm}
\section{Introduction}
\label{sec:intro}
\vspace{-1mm}
Many approaches have been researched in the field of music separation such as local Gaussian modelling \cite{mss_gmodel_1, mss_gmodel_2}, non-negative matrix factorization \cite{mss_nmf_1, mss_nmf_2, mss_nmf_3}, kernel additive modelling \cite{mss_kernel_additive_1} and hybrid methods combining these approaches \cite{mss_combi_1, mss_combi_2}.
In particular, many methods have been investigated which introduce \emph{deep neural networks} (DNNs) in order to improve the separation performance in recent years.
There are three basic DNN architectures, namely \emph{multi-layer perceptrons} (MLPs) \cite{mlp_org}, \emph{convolutional neural networks} (CNNs) \cite{cnn_org} and \emph{recurrent neural networks} (RNNs) \cite{rnn_org}, and all of them have been already introduced for the task of audio source separation.
For instance, an MLP was used to separate the input spectra and then obtain separated results in \cite{mss_mlp_1, mss_mlp_2}.
In \cite{mss_cnn_1, mss_rnn_1}, CNNs and RNNs were used to realize source separation with improved quality than previous MLP-based methods since CNNs and RNNs can consider the temporal contexts via convolution and recurrent layers. 

Although the aforementioned literature has improved the performance of music separation drastically, there are two problems with respect to training the separation networks: (1) most existing methods consider only the time or the frequency domain but not both, and, (2) they do not consider the mutual influence among output sources since loss functions are independently applied to each estimated source and the corresponding ground truth.
For example, one of the state-of-the-art open-source systems for music source separation, called \emph{Open-Unmix} (UMX) \cite{umx}\footnote{\url{https://github.com/sigsep/open-unmix-[nnabla|pytorch]}}, conducts music source separation only in the frequency domain as the input and output of UMX are both spectrograms.
Furthermore, UMX applies the conventional \emph{mean squared error} (MSE) loss function to individual pairs of estimated and corresponding ground truth spectrograms for each instrument.
In other words, UMX trains networks one-by-one for each instrument.

In order to solve these problems, we propose schemes with respect to the used loss function and model architecture.
In the field of speech enhancement, which is a special case of audio source separation, methods considering time and frequency domain have been researched in recent years \cite{mdphd, hifi_gan}.
For instance, Kim \textit{et al.} showed in \cite{mdphd} the effectiveness of multi-domain processing via hybrid denoising networks.
Furthermore, Su \textit{et al.} reported in \cite{hifi_gan} that building two discriminators which are responsible for time and frequency domain can realize effective denoising and dereverberation in their scheme of using \emph{generative adversarial networks} (GANs).
Inspired by these reports, we believe that considering both, time and frequency domain, is important to realize effective music source separation.
Next, in the field of audio source separation, the effectiveness of Conv-TasNet~\cite{conv_tasnet, conv_tasnet_mss}, a fully-convolutional time-domain audio separation network, was reported.
In particular, D\'{e}fossez \textit{et al.} reported in \cite{conv_tasnet_mss} that the performance of Conv-TasNet was higher than the one of UMX when trained and evaluated on the same dataset.
One of the reasons in our opinion is that the architecture of Conv-TasNet allows information sharing among sources as the convolutional layers consider all input channels when computing one output channel. 
In contrast, UMX does not allow such information sharing as it trains independent source extraction networks for each instrument.
Therefore, it is difficult for UMX to consider the mutual influence among instruments obtained from the same input mixture.

Motivated by this discussion, we propose two new loss functions and a new model architecture which we add to UMX, called \emph{CrossNet-UMX} (X-UMX)\footnote{Our implementation for NNabla, Sony's deep learning framework, is available at \url{https://github.com/sony/ai-research-code/tree/master/x-umx}. 
Furthermore, a PyTorch version building upon Asteroid \cite{asteroid} is available at \url{https://github.com/asteroid-team/asteroid/tree/master/egs/musdb18/X-UMX}.}.
First, we introduce a \emph{multi-domain loss} (MDL) where we append an additional differentiable \emph{short-time Fourier transform} (STFT) or \emph{inverse STFT} (ISTFT) layer\footnote{If the network outputs a spectrogram, we append an ISTFT layer whereas a STFT layer is added if the network output is a time signal.} during training only. The loss is computed from the estimates before and after the STFT/ISTFT layer.
In this way, MDL can consider not only frequency but also time domain differences between estimates and ground truth.
Second, we also propose a further loss function, named \emph{combination loss} (CL), and bridging network paths for UMX.
As we mentioned above, not only UMX but also almost all conventional methods for music source separation train their networks for each source independently.
Thus, it is difficult to find the root-cause of performance degradation, i.e., which sources are leaking and thus producing incorrect instrument estimates.
To tackle this problem, CL considers the relationship among output sources by also producing output spectrograms for instrument combinations and applying MDL to them. 
If the performance of the $i$th source separation is insufficient, other combinations including the $i$th source will be adversely affected while the others which do not include the $i$th source are not.
In addition, we bridge the network paths of UMX for different instruments and thus share information among all instruments. Our bridging operation is beneficial for networks like UMX which consists of individual extraction networks and not one joint separation network.
Hence, a network like Conv-TasNet which is already crossing among sources via convolutional layers does not need this operation.

The above proposed loss functions, i.e., MDL and CL, only affect the training step and, therefore, can be introduced to many conventional methods since they are merely loss functions.
In addition, our bridging operation requires only a slight network modification but does not increase the number of parameters that need to be learned.
Consequently, performance improvements can be gained for most DNN-based source separation methods with introducing almost no additional computational costs at inference time.

\vspace{-1mm}
\section{Proposed Loss Functions}
\label{sec:propose}
\vspace{-1mm}
In this section, we describe in detail our new loss functions, i.e., MDL and CL, and discuss their merits.
We assume that the time-domain mixture signal $\bm{x}$ consists of $J$ sources, i.e.,
\begin{align}
\bm{x} = \sum_{j=1}^{J} \bm{y}_j,
\label{eq:1}
\end{align} 
where $\bm{y}_j$ denotes the time-domain signal of the $j$th source and $\bm{x}$, $\bm{y}_j$ are column vectors with the samples.
Furthermore, we assume that the output of the DNN is a mask $\bm{M}_j$ which can extract the $j$th desired source from the mixture spectrum $\bm{X} = \mathcal{S} \{ \bm{x} \}$:
\begin{align}
\hat{\bm{y}}_j &= \mathcal{S}^{-1} \{ \hat{\bm{Y}}_j \} = \mathcal{S}^{-1} \{ \bm{M}_j \circ \bm{X} \},
\label{eq:2}
\end{align} 
where $\mathcal{S}$ and $\mathcal{S}^{-1}$ are the forward and inverse operators of the STFT, respectively.
Furthermore, $\hat{\bm{y}}_j$ and $\hat{\bm{Y}}_j$ are the predicted results of time and frequency domain ground truths $\bm{y}_j$ and $\bm{Y}_j$.

\vspace{-1mm}
\subsection{Multi-Domain Loss (MDL)}
\label{subsec:MDL}
\vspace{-1mm}
\begin{figure}[!t]
\begin{minipage}[b]{1.0\linewidth}
  \centering
  \centerline{\includegraphics[width=8.75cm]{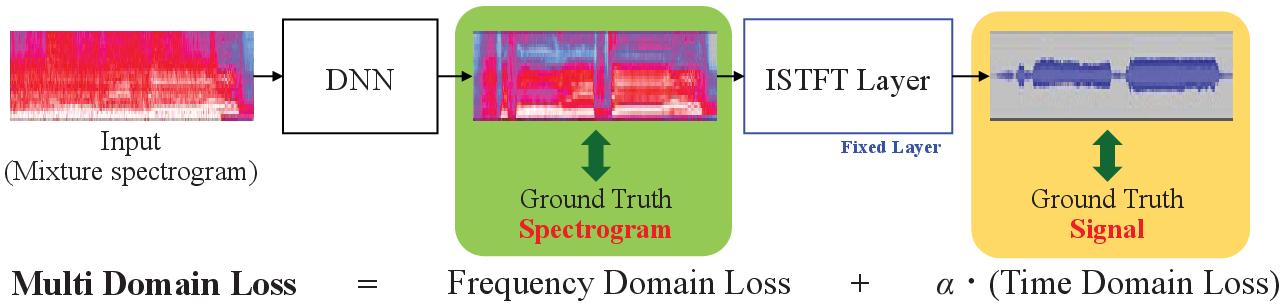}}
  \flushleft{(a) Frequency domain network with appended ISTFT layer.}\medskip
\end{minipage}
\begin{minipage}[b]{1.0\linewidth}
\vspace{+1mm}
  \centering
  \centerline{\includegraphics[width=8.75cm]{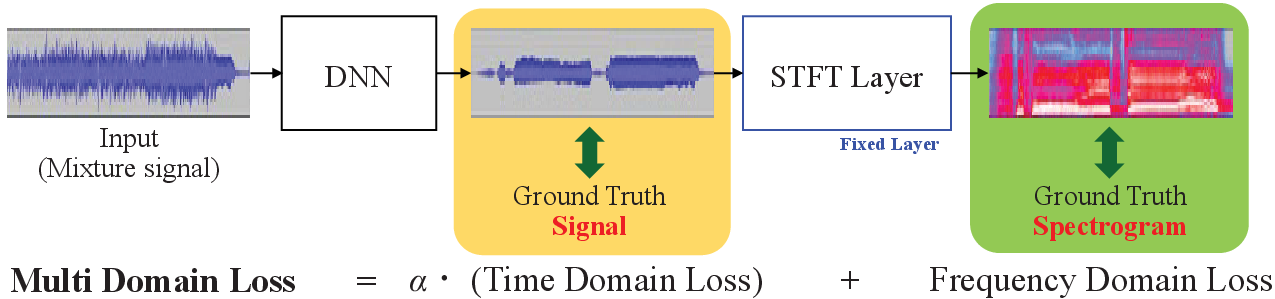}}
  \flushleft{(b) Time domain network with appended STFT layer.}
\end{minipage}
\vspace{-3mm}
\caption{Multi-domain loss (MDL). Note that $\alpha$ is a scaling parameter.}
\label{fig:MDL}
\vspace{-3mm}
\end{figure}
%
 \begin{figure}[!t]
  \begin{center}
\includegraphics[width=8.75cm]{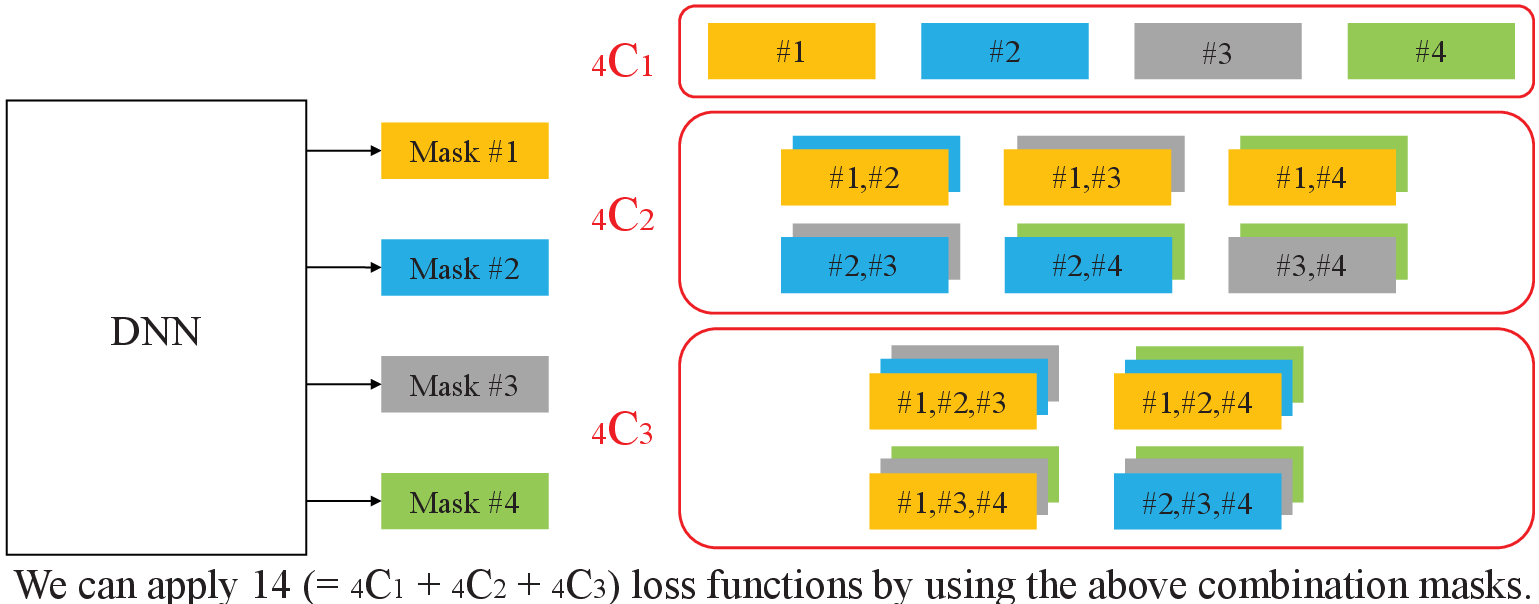}
  \caption{Combination Loss (CL) for the case that the mixture consists of four sources.}
   \label{fig:CL}
   \vspace{-8mm}
  \end{center}
\end{figure}
%
In the scheme of MDL, we first append an additional differentiable and fixed STFT or ISTFT layer after the output layer as shown in Fig.~\ref{fig:MDL}.
This allows us to compute the loss in the time as well as the frequency domain.
Note that the STFT or ISTFT layer does not affect the inference step since this layer is only used during training for computing the MDL.
In our method, we use the \emph{mean squared error} (MSE) between separated and ground truth spectrograms as frequency domain loss, and the \emph{weighted signal-to-distortion ratio} (wSDR) \cite{dcunet_1} as time domain loss, i.e.,
\begin{align}
\mathcal{L}_{\mbox{\tiny{MDL}}}^J = \mathcal{L}_{\mbox{\tiny{MSE}}}^J + \alpha \mathcal{L}_{\mbox{\tiny{wSDR}}}^J,
\label{eq:3}
\end{align}
where $\alpha$ is a scaling parameter for mixing multiple domains of loss.
$\mathcal{L}_{\mbox{\tiny{MSE}}}^J$ and $\mathcal{L}_{\mbox{\tiny{wSDR}}}^J$ are respectively calculated as follows:
\begin{subequations}
\begin{align}
\mathcal{L}_{\mbox{\tiny{MSE}}}^J &= 
    \sum_{j=1}^J \sum_{t,f} \left\{\lvert Y_j (t, f)\rvert - \lvert\hat{Y}_j (t, f)\rvert\right\}^2, \\
\mathcal{L}_{\mbox{\tiny{wSDR}}}^J &= 
    \sum_{j=1}^J \left\{-\rho_j \frac{\bm{y}_j^\mathrm{T}\hat{\bm{y}}_j}{\|\bm{y}_j\| \: \|\hat{\bm{y}}_j\|} 
    - (1-\rho_j) \frac{(\bm{x}-\bm{y}_j)^\mathrm{T}(\bm{x}-\hat{\bm{y}}_j)}{\|\bm{x}-\bm{y}_j\| \: \|\bm{x}-\hat{\bm{y}}_j\|} \right\}, 
\label{eq:4}
\end{align}
\end{subequations}
where $j$ denotes the $j$th source.
Furthermore, $t$ and $f$ denote the frame and frequency bin index of the spectrogram $Y_j(t,f)$ and its estimate $\hat{Y}_j (t, f)$, respectively.
In addition, $\rho_j$ is the energy ratio between the $j$th source $\bm{y}_j$ and the mixture $\bm{x}$ in the time-domain, i.e., $\rho_j = \|\bm{y}_j\|^2 / (\|\bm{y}_j\|^2 + \|\bm{x} - \bm{y}_j\|^2)$.
Please note that the output range of wSDR in Eq.~\eqref{eq:4} is bounded to $[-1, 1]$. This range is useful for a stable training and no ``log'' operation is required (although SDR is traditionally calculated including the logarithm).

Using MDL, we take advantage of both domains, which provides a performance improvement as we will see in Sec.~\ref{sec:exp}. 
As MDL is merely a loss function, it can be used for many conventional methods without requiring additional calculations during inference. 
Please note that we focus in this paper on frequency domain networks, e.g., UMX, as this approach is more common than time domain networks.

\begin{figure*}[t]
\centering
\subfloat[UMX]{
\includegraphics[scale=0.47]{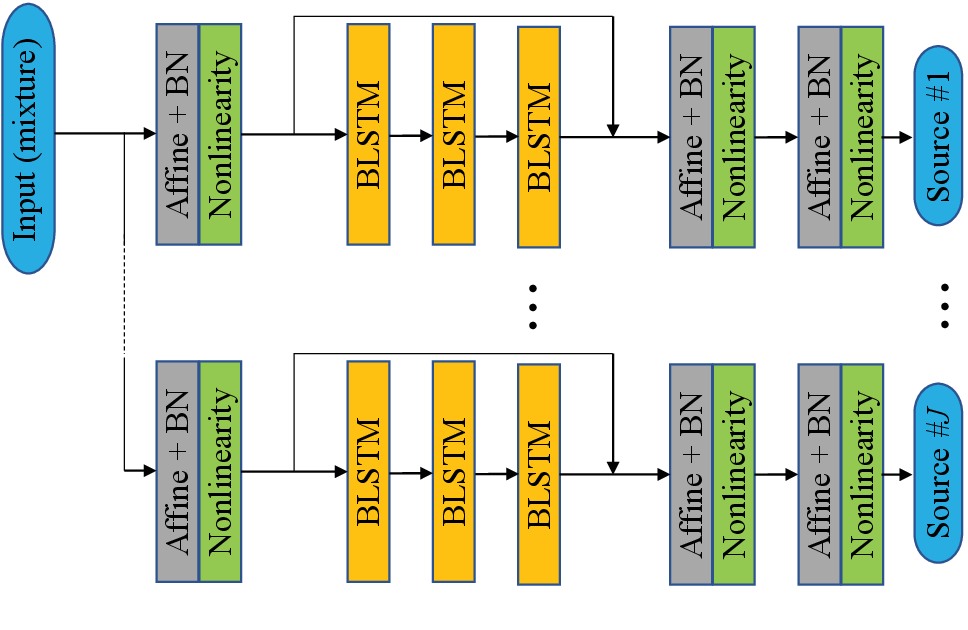}
\label{fig:umx}}
\hspace{+3mm}
\subfloat[X-UMX]{
\includegraphics[scale=0.47]{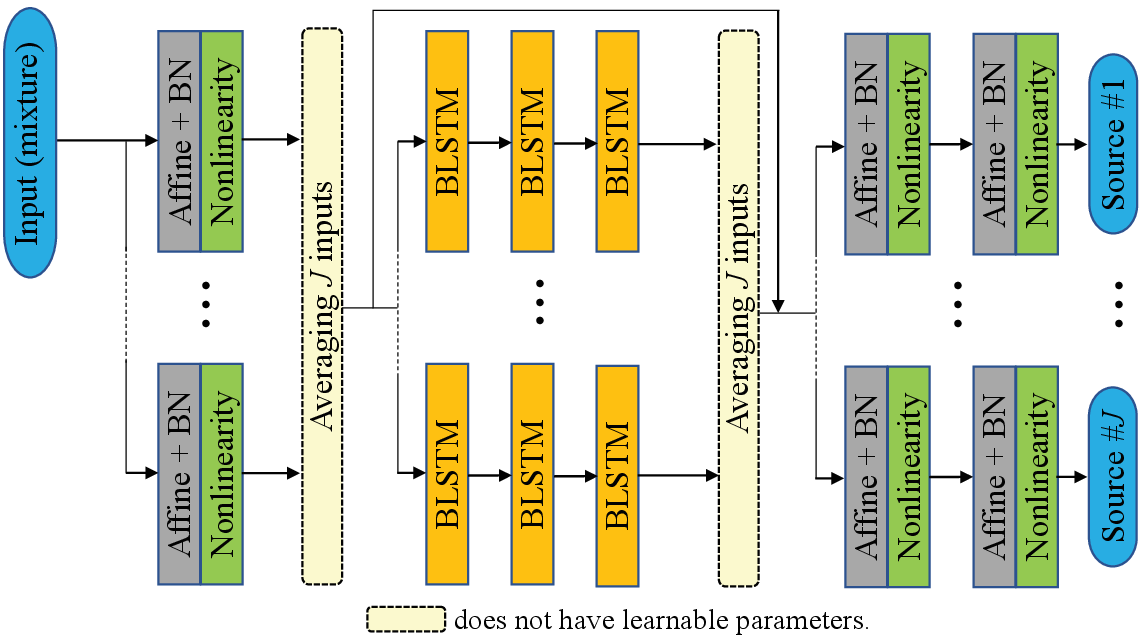}
\label{fig:cross}}

  \caption{Comparison of network architectures used in our experiments.}
  \label{fig:comp_archi}
  \vspace{-3mm}
\end{figure*}

\vspace{-1mm}
\subsection{Combination Schemes}
\label{subsec:combi}
\vspace{-1mm}
In this subsection, we introduce the new combination loss (Sec.~2.2.1) and our new bridging network architecture (Sec.~2.2.2) to help each UMX's extraction network to support each other.

\vspace{-1mm}
\subsubsection{Combination Loss (CL)}
\label{subsec:CL}
\vspace{-1mm}
In the scheme of CL, we consider the combinations of output masks.
Specifically, we combine two or more estimated masks into new ones where each of them can extract two or more sources from the mixture.
By using the newly obtained combination masks, we can compute more loss functions than if we only compare each estimated mask with its target, i.e.,
\begin{align}
\mathcal{L} = \frac{1}{N} \sum_{n=1}^N  \mathcal{L}_{\mbox{\tiny{MDL}}}^n,
\label{eq:5}
\end{align}
where $N > J$ is the total number of possible combinations except for mixing all $J$ sources, namely $N = \sum_{i=1}^{J-1} {}_J\mathrm{C}_i$, and $n$ denotes the index of $n$th combination, where ${}_J\mathrm{C}_i$ denote the number of possible combinations of $i\:(=1,2,\cdots,J-1)$ instruments from the total of $J$ instruments in the mixture, i.e., ${}_J\mathrm{C}_i$ is equal to the binomial coefficient $\binom{J}{i}$.
For example, in the situation of separating four sources, we can consider 14 $(={}_4\mathrm{C}_1 + {}_4\mathrm{C}_2 + {}_4\mathrm{C}_3)$ combinations in total as shown in Fig.~\ref{fig:CL}, 
while conventional methods consider each source independently\footnote{Initial experiments showed that the combination ${}_J\mathrm{C}_J$, i.e., the case of $i=J$, is not adding further performance improvements and, hence, it is not used in Eq.~\eqref{eq:5}.}.

In order to illustrate the benefit of CL, let us consider the following example: Assume that we have a system with leakage of \emph{vocals} into \emph{drums} and \emph{other} resulting in similar errors that both instruments exhibit. 
By considering the combination \emph{drums + other}, we will notice that the two errors are correlated, resulting in an even larger leakage of vocals which we try to mitigate by using our proposed CL loss.

\vspace{-1mm}
\subsubsection{Bridging Networks}
\label{subsec:bridging}
\vspace{-1mm}
In our method, CL aims to train each network with considering the relationship among output sources by combining output masks.
In addition, we observed that it is effective to cross not only the loss function via CL but also the network graphs in order to help each UMX's extraction network to support each other.
Hence, we also propose UMX with a crossing architecture, named \emph{CrossNet-UMX} (X-UMX).
Specifically, we connect the paths to cross each source's network by adding just two average operators to the original UMX model as shown in Fig.~\ref{fig:comp_archi}. 
Using these average operations does not change the total numbers of parameters which would not be the case if we would use a concatenation operation.
Please note that this average operation is only needed for models like UMX since UMX consists of individual extraction networks.

In this way, our method can consider multiple sources together, i.e., two or more source separation, than considering each source independently.
From a different viewpoint, CL particularly can be considered to provide a benefit similar to multi-task learning due to considering multiple sources jointly by computing combinational masks.
\\

We will see in Sec.~\ref{sec:exp} that MDL and our combination schemes provide a performance improvement for UMX. We can expect such an improvement also for many conventional methods without having to introduce additional computational costs at inference time since MDL and CL are merely loss functions and the bridging is achieved with a simple average operation without learnable parameters.

 \begin{figure*}[!thb]
\centering
\subfloat[SDR]{
\includegraphics[scale=0.4825]{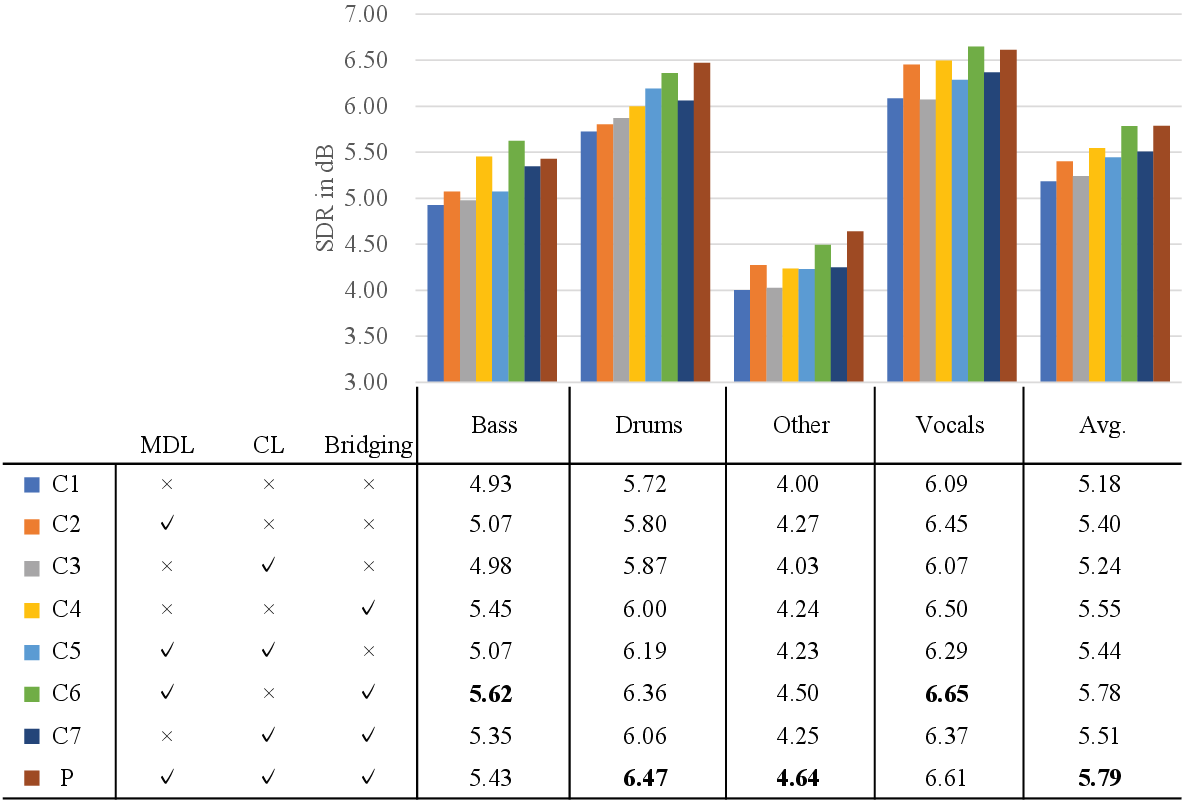}
\label{fig:result_sdr}}
\subfloat[SAR]{
\includegraphics[scale=0.53]{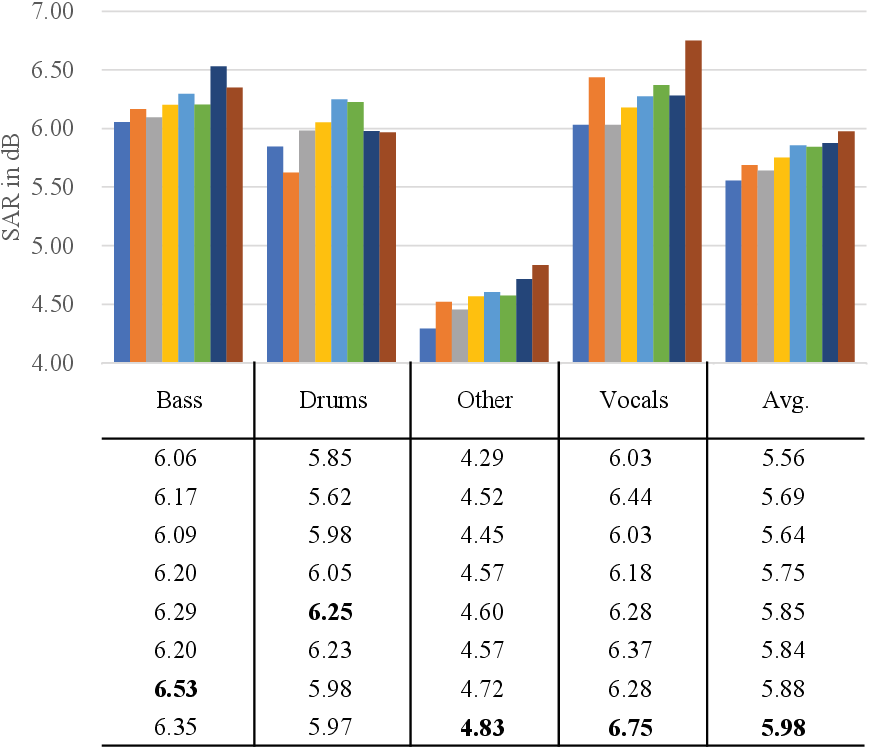}
\label{fig:result_sar}}
\vspace{-1mm}
\caption{Experimental results for proposed methods.}
\label{fig:results}
\vspace{-2mm}
\end{figure*}

\vspace{-1mm}
\section{Experiments}
\label{sec:exp}
\vspace{-1mm}
In this section, we conduct music separation experiments in order to confirm the validity of our method.
The task is to separate a song into its four constituent instruments.

\subsection{Setup}
\label{subsec:setup}
In our experiments, we evaluate the proposed method on the MUSDB18 dataset \cite{musdb18} which is comprised of 150 songs each of which is recorded at 44.1kHz sampling rate.
MUSDB18 consists of two subsets (`train' and `test') where we split the train set further into `train' and `valid' as defined in the `musdb' package\footnote{\url{https://github.com/sigsep/sigsep-mus-db/blob/master/musdb/configs/mus.yaml}}. 
For each song, the mixture and its four sources, i.e., \textit{bass, drums, other} and \textit{vocals}, are available.
As in the original UMX, we operated our networks in the STFT magnitude domain using a Hann window of length 4096 with 75\% window overlap.

Since CL needs to be applied to the joint instrument network due to the usage of the combinations of the output masks, we cannot use the original UMX implementation, which independently builds and trains a network for each instrument.
Hence, we always train in the following the four separation networks jointly, even in the case that no combination scheme is used, and the loss function is merely the mean of the four individual losses for each instrument.
This has the effect that the early stopping at the epoch with the smallest validation error is not done for each instrument individually (as is the case for the original UMX) but the early stopping is done at the same epoch for all four networks.
Furthermore, the learning rate drops, which are determined by the `ReduceLROnPlateau` function are done at the same epoch.
Hence, the results that we obtain in Fig.~\ref{fig:results} for ``C1'' differ from the results of the original UMX.
Besides the modifications mentioned above, all other experimental settings are the same as in the original UMX.

Finally, please note that the scaling parameter $\alpha$, introduced in Eq.~\eqref{eq:3} for MDL, was set to $\alpha = 10$ in order to approximately equalize the ranges of $\mathcal{L}^J_{\mbox{\tiny MSE}}$ and $\mathcal{L}^J_{\mbox{\tiny wSDR}}$ by looking at the individual loss function's learning curves.

\vspace{-1mm}
\subsection{Results}
\label{subsec:results}
\vspace{-1mm}
To evaluate the performance of our method, we used the \emph{signal-to-distortion ratio} (SDR) and \emph{sources-to-artifacts ratio} (SAR) computed with BSSEval v4 called `museval'\footnote{\url{https://github.com/sigsep/sigsep-mus-eval}}, which was also the official evaluation scheme for SiSEC 2018~\cite{stoter20182018}.
\begin{table}[tb]
\centering
\caption{Comparison of X-UMX with other public methods in terms of SDR (``median of frames, median of tracks'').}
\vspace{-2mm}
\resizebox{\linewidth}{!}{
\begin{tabular}{ l | c c c c c }
	\hline
    \textbf{Method} 
                 & \textbf{Bass} & \textbf{Drums} & \textbf{Other} & \textbf{Vocals} & \textbf{Avg.}\\ \hline \hline
UMX \cite{umx} & 5.07 & 6.04 & 4.28 & 6.25 & 5.41 \\
Meta-TasNet \cite{meta_tasnet} & 5.58 & 5.91 & 4.19 & 6.40 & 5.52 \\
Demucs \cite{demucs} & \textbf{5.83} & 6.08 & 4.12 & 6.29 & 5.58 \\ 
Conv-TasNet \cite{conv_tasnet_mss} & 5.66 & 6.08 & 4.37 & \textbf{6.81} & 5.73 \\
X-UMX (proposed) & 5.43 & \textbf{6.47} & \textbf{4.64} & 6.61 & \textbf{5.79} \\
\hline
\end{tabular}
\label{tb:methods_pub}
}
\vspace{-3.5mm}
\end{table}

The experimental results are shown in Fig.~\ref{fig:results}.
Note that C1 is almost identical to original UMX since it uses the same network architecture and loss function.
However, C1's performance is inferior compared to the original UMX.
As we discussed in Sec.~\ref{subsec:setup}, this is due to the difference in the early stopping and learning rate drops.
First, each of our contributions, i.e., MDL, CL and Bridging, respectively enable UMX to improve performance since the results of C2-C4 outperform those of C1.
Next, we can confirm that all methods that use two modifications compared to the original UMX, i.e., C5-C7, also outperformed C1.
Particularly, collaborative using MDL and network bridging (C6) improved the SDR values of C1 drastically, which increases the average score from $5.18$dB to $5.78$dB.
We can observe that the improvements of MDL and bridging linearly add up by confirming the differences of the average score as follows:
$\mbox{C2} - \mbox{C1} = 0.22$dB, $\mbox{C4} - \mbox{C1} =0.37$dB. 
Then the sum of them, i.e., $0.22$dB + $0.37$dB $=0.59$dB, is nearly equal to the improvement from C1 to C6 ($=0.60$dB).
Meanwhile, we can confirm that only CL seems to be not always provide improvement in terms of SDR when it is used with another scheme by focusing on the values of C4 (w/ Bridging) and C7 (w/ Bridging and CL).
On the other hand, we can observe that SAR is always improved by adding CL as shown in Fig.~\ref{fig:results}(b).
In particular, the SARs were improved as follows: +0.08 (C1 $\rightarrow$ C3), +0.16 (C2 $\rightarrow$ C5), +0.13 (C4 $\rightarrow$ C7) and +0.14 (C6 $\rightarrow$ P).
Thus, CL is effective to reduce artifacts and it is beneficial to introduce it since the improved SAR score leads to an improved sound quality as shown in \cite{ward2018bss}.
Finally, we can confirm that using all our proposed modifications jointly, i.e., MDL, CL and bridging operation, which is denoted as `P' in Fig.~\ref{fig:results}, gives the best performance in terms of SDR and SAR among all methods.
In addition, we can compare our method to other music source separation public state-of-the-art systems shown in Table~\ref{tb:methods_pub}.
\\

\vspace{-2mm}
\section{Conclusions}
\label{sec:concludions}
\vspace{-1mm}
In this paper, we proposed two new loss functions called \emph{multi-domain loss} (MDL) and \emph{combination loss} (CL) and a suitable network architecture called \emph{CrossNet-UMX} (X-UMX) realized by a bridging operation.
We showed that MDL and CL are effective and convenient for music separation methods working in the frequency domain since both are loss functions which are used during training and thus do not change the inference step.
Hence, it is easy to apply them to many other conventional DNN-based methods.
In addition, if the target network is built up from sub-networks for each source, we showed that bridging the network paths by a simple averaging operation helps each extraction network to know each other better and, thus, improve performance.
In this paper, we applied MDL and CL to a well-known and state-of-the-art open source library Open-Unmix (UMX), by only adding two average operators to the X-UMX model and the improved results showed the validity of our approach. 

\clearpage
\bibliographystyle{IEEEbib}
\bibliography{refs_euc}

\begin{thebibliography}{10}

\bibitem{mss_gmodel_1}
N.~Q.~K. {Duong}, E.~{Vincent}, and R.~{Gribonval},
\newblock ``Under-determined reverberant audio source separation using a
  full-rank spatial covariance model,''
\newblock {\em IEEE Trans. on Audio, Speech, and Language Processing}, vol. 18,
  no. 7, pp. 1830--1840, 2010.

\bibitem{mss_gmodel_2}
D.~{FitzGerald}, A.~{Liutkus}, and R.~{Badeau},
\newblock ``{PROJET} --- {S}patial audio separation using projections,''
\newblock in {\em Proc. of IEEE International Conference on Acoustics, Speech
  and Signal Processing (ICASSP)}, 2016, pp. 36--40.

\bibitem{mss_nmf_1}
A.~{Liutkus}, D.~{Fitzgerald}, and R.~{Badeau},
\newblock ``Cauchy nonnegative matrix factorization,''
\newblock in {\em Proc. of IEEE Workshop on Applications of Signal Processing
  to Audio and Acoustics (WASPAA)}, 2015, pp. 1--5.

\bibitem{mss_nmf_2}
J.~{Le Roux}, J.~R. {Hershey}, and F.~{Weninger},
\newblock ``Deep nmf for speech separation,''
\newblock in {\em Proc. of IEEE International Conference on Acoustics, Speech
  and Signal Processing (ICASSP)}, 2015, pp. 66--70.

\bibitem{mss_nmf_3}
Y.~{Mitsufuji}, S.~{Koyama}, and H.~{Saruwatari},
\newblock ``Multichannel blind source separation based on non-negative tensor
  factorization in wavenumber domain,''
\newblock in {\em Proc. of IEEE International Conference on Acoustics, Speech
  and Signal Processing (ICASSP)}, 2016, pp. 56--60.

\bibitem{mss_kernel_additive_1}
A.~{Liutkus}, D.~{Fitzgerald}, Z.~{Rafii}, B.~{Pardo}, and L.~{Daudet},
\newblock ``Kernel additive models for source separation,''
\newblock {\em IEEE Trans. on Signal Processing}, vol. 62, no. 16, pp.
  4298--4310, 2014.

\bibitem{mss_combi_1}
A.~{Ozerov} and C.~{Fevotte},
\newblock ``Multichannel nonnegative matrix factorization in convolutive
  mixtures for audio source separation,''
\newblock {\em IEEE Trans. on Audio, Speech, and Language Processing}, vol. 18,
  no. 3, pp. 550--563, 2010.

\bibitem{mss_combi_2}
A.~{Liutkus}, D.~{Fitzgerald}, and Z.~{Rafii},
\newblock ``Scalable audio separation with light kernel additive modelling,''
\newblock in {\em Proc. of IEEE International Conference on Acoustics, Speech
  and Signal Processing (ICASSP)}, 2015, pp. 76--80.

\bibitem{mlp_org}
F.~Rosenblatt,
\newblock {\em Principles of Neurodynamics: Perceptrons and the Theory of Brain
  Mechanisms},
\newblock Defense Technical Information Center, 1961.

\bibitem{cnn_org}
K.~Fukushima,
\newblock ``{N}eocognitron: {A} self-organizing neural network model for a
  mechanism of pattern recognition unaffected by shift in position,''
\newblock {\em Biological Cybernetics}, vol. 36, pp. 193--202, 1980.

\bibitem{rnn_org}
D.~E. Rumelhart, G.~E. Hinton, and R.~J. Williams,
\newblock ``Learning representations by back-propagating errors,''
\newblock {\em Nature}, vol. 323, no. 6088, pp. 533--536, 1986.

\bibitem{mss_mlp_1}
A.~A. {Nugraha}, A.~{Liutkus}, and E.~{Vincent},
\newblock ``Multichannel music separation with deep neural networks,''
\newblock in {\em Proc. of 24th European Signal Processing Conference
  (EUSIPCO)}, 2016, pp. 1748--1752.

\bibitem{mss_mlp_2}
S.~{Uhlich}, F.~{Giron}, and Y.~{Mitsufuji},
\newblock ``Deep neural network based instrument extraction from music,''
\newblock in {\em Proc. of IEEE International Conference on Acoustics, Speech
  and Signal Processing (ICASSP)}, 2015, pp. 2135--2139.

\bibitem{mss_cnn_1}
N.~{Takahashi} and Y.~{Mitsufuji},
\newblock ``Multi-scale multi-band densenets for audio source separation,''
\newblock in {\em Proc. of IEEE Workshop on Applications of Signal Processing
  to Audio and Acoustics (WASPAA)}, 2017, pp. 21--25.

\bibitem{mss_rnn_1}
S.~{Uhlich}, M.~{Porcu}, F.~{Giron}, M.~{Enenkl}, T.~{Kemp}, N.~{Takahashi},
  and Y.~{Mitsufuji},
\newblock ``Improving music source separation based on deep neural networks
  through data augmentation and network blending,''
\newblock in {\em Proc. of IEEE International Conference on Acoustics, Speech
  and Signal Processing (ICASSP)}, 2017, pp. 261--265.

\bibitem{umx}
F.-R. St\"{o}ter, S.~Uhlich, A.~Liutkus, and Y.~Mitsufuji,
\newblock ``Open-{U}nmix - {A} reference implementation for music source
  separation,''
\newblock {\em Journal of Open Source Software}, vol. 4, pp. 1667, 09 2019.

\bibitem{mdphd}
J.-H. Kim, J.~Yoo, S.~Chun, A.~Kim, and J.-W. Ha,
\newblock ``Multi-domain processing via hybrid denoising networks for speech
  enhancement,''
\newblock {\em arXiv}, 2018.

\bibitem{hifi_gan}
J.~Su, Z.~Jin, and A.~Finkelstein,
\newblock ``{HiFi-GAN}: {H}igh-fidelity denoising and dereverberation based on
  speech deep features in adversarial networks,''
\newblock in {\em Proc. of Interspeech}, 2020 (accepted for publication).

\bibitem{conv_tasnet}
Y.~{Luo} and N.~{Mesgarani},
\newblock ``{Conv-TasNet}: {S}urpassing ideal time-frequency magnitude masking
  for speech separation,''
\newblock {\em IEEE/ACM Trans. on Audio, Speech, and Language Processing}, vol.
  27, no. 8, pp. 1256--1266, 2019.

\bibitem{conv_tasnet_mss}
A.~D\'{e}fossez, N.~Usunier, L.~Bottou, and F.~Bach,
\newblock ``Music source separation in the waveform domain,''
\newblock {\em arXiv}, 2019.

\bibitem{asteroid}
M.~Pariente, S.~Cornell, J.~Cosentino, S.~Sivasankaran, E.~Tzinis,
  J.~Heitkaemper, M.~Olvera, F.-R. Stöter, M.~Hu, J.~M. Martín-Doñas,
  D.~Ditter, A.~Frank, A.~Deleforge, and E.~Vincent,
\newblock ``Asteroid: the {PyTorch}-based audio source separation toolkit for
  researchers,''
\newblock in {\em Proc. of Interspeech}, 2020.

\bibitem{dcunet_1}
H.-S. Choi, J.-H. Kim, J.~Huh, A.~Kim, J.-W. Ha, and K.~Lee,
\newblock ``Phase-aware speech enhancement with deep complex {U}-net,''
\newblock {\em arXiv}, 2019.

\bibitem{musdb18}
Z.~Rafii, A.~Liutkus, F.-R. St{\"o}ter, S.~I. Mimilakis, and R.~Bittner,
\newblock ``The {MUSDB18} corpus for music separation,'' Dec. 2017.

\bibitem{stoter20182018}
F.-R. St{\"o}ter, A.~Liutkus, and N.~Ito,
\newblock ``The 2018 signal separation evaluation campaign,''
\newblock in {\em International Conference on Latent Variable Analysis and
  Signal Separation}. Springer, 2018, pp. 293--305.

\bibitem{meta_tasnet}
D.~{Samuel}, A.~{Ganeshan}, and J.~{Naradowsky},
\newblock ``Meta-learning extractors for music source separation,''
\newblock in {\em Proc. of IEEE International Conference on Acoustics, Speech
  and Signal Processing (ICASSP)}, 2020, pp. 816--820.

\bibitem{demucs}
A.~D\'{e}fossez, N.~Usunier, L.~Bottou, and F.~Bach,
\newblock ``Demucs: {D}eep extractor for music sources with extra unlabeled
  data remixed,''
\newblock {\em arXiv}, 2019.

\bibitem{ward2018bss}
D.~Ward, H.~Wierstorf, R.~D. Mason, E.~M. Grais, and M.~D. Plumbley,
\newblock ``Bss eval or peass? predicting the perception of singing-voice
  separation,''
\newblock in {\em 2018 IEEE International Conference on Acoustics, Speech and
  Signal Processing (ICASSP)}, 2018, pp. 596--600.

\end{thebibliography}

\end{document}